\documentstyle[12pt,epsfig]{article}
\renewcommand{\theequation}{\mbox{$\thesection . \arabic{equation}$}}

\begin{document}
\newcommand{\beq}{\begin{equation}}\newcommand{\eeq}{\end{equation}}
\newcommand{\barr}{\begin{eqnarray}}\newcommand{\earr}{\end{eqnarray}}

\newcommand{\andy}[1]{ }

\def\txt{\textstyle}
\def\NI{\noindent}   \def\lslittlenote#1{\lsnote{{\sl #1}}}
 \def\lsnote#1{\def\dash{\hbox{\rm---}}{\bf~[[}~{\tt #1}~{\bf]]~}}
\def\eqn#1{Eq.\ (\ref{eq:#1})} \def\pp{{\vec p}}
\def\ask{\marginpar{?? ask:  \hfill}}  \def\fin{\marginpar{fill in ...
\hfill}}
\def\note{\marginpar{note \hfill}}  \def\check{\marginpar{check \hfill}}
\def\discuss{\marginpar{discuss \hfill}}
\def\hh{\widehat}
\def\wtilde{\widetilde}  
\newcommand{\bm}[1]{\mbox{\boldmath $#1$}}
\newcommand{\bmsub}[1]{\mbox{\boldmath\scriptsize $#1$}}
\newcommand{\bmh}[1]{\mbox{\boldmath $\hat{#1}$}}
\newcommand{\ket}[1]{| #1 \rangle}
\newcommand{\bra}[1]{\langle #1 |}

\begin{titlepage}
\begin{flushright}
\today \\
BA-TH/99-359\\
\end{flushright}
\vspace{.5cm}
\begin{center}
{\LARGE Spontaneous emission and lifetime modification caused by an
intense electromagnetic field

}

\quad

{\large P. FACCHI and S. PASCAZIO \\
           \quad    \\

        Dipartimento di Fisica, Universit\`a di Bari

     and Istituto Nazionale di Fisica Nucleare, Sezione di Bari \\
 I-70126 Bari, Italy

}

\vspace*{.5cm} PACS: 42.50.Hz; 42.50.Vk; 03.65.Bz
\vspace*{.5cm}

{\small\bf Abstract}\\ \end{center}

{\small We study the temporal evolution of a three-level system (such
as an atom or a molecule), initially prepared in an excited state,
bathed in a laser field tuned at the transition frequency of the
other level. The features of the spontaneous emission are
investigated and the lifetime of the initial state is evaluated: a
Fermi ``golden rule" still applies, but the on-shell matrix elements
depend on the intensity of the laser field. In general, the lifetime
is a decreasing function of the laser intensity. The phenomenon we
discuss can be viewed as an ``inverse" quantum Zeno effect and can be
analyzed in terms of dressed states.

}

\end{titlepage}

\newpage

\setcounter{equation}{0}
\section{Introduction }
 \label{sec-introd}
 \andy{intro}

The temporal behavior of quantum mechanical systems can be strongly
influenced by the action of an external agent. A good example is the
quantum Zeno effect \cite{shortt,QZE}, where the quantum mechanical
evolution of a given (not necessarily unstable) state is slowed down
(or even halted) by performing a series of measurements that
ascertain whether the system is still in its initial state. This
peculiar effect is historically associated and usually ascribed to
what we could call a ``pulsed" quantum mechanical observation on the
system. However, it can also be obtained by performing a
``continuous" observation of the quantum state, e.g.\ by means of an
intense field \cite{MPS,contvspulsed}.

Most experiment that have been performed or proposed in order to
modify the quantum mechanical evolution law make use of oscillating
systems \cite{Cook,Itano,PNBR,Zeilinger,PL}. On the other hand, it
would be interesting to understand whether and to which extent the
evolution law of a {\em bona fide} ``unstable" system can be
changed. In order to discuss the evolution of genuine unstable
systems one usually makes use of the Weisskopf-Wigner approximation
\cite{seminal}, that ascribes the main properties of the decay law
to a pole located near the real axis of the complex energy plane.
This yields the Fermi ``golden rule" \cite{Fermigold}. In this
paper we shall investigate the possibility that the lifetime of an
unstable quantum system can be modified by the presence of a very
intense electromagnetic field. We shall look at the temporal
behavior of a three-level system (such as an atom or a molecule),
where level \#1 is the ground state and levels \#2, \#3 are two
excited states. (See Figure \ref{fig:fig0}.) The system is
initially prepared in level \#2 and if it follows its natural
evolution, it will decay to level \#1. The decay will be
(approximately) exponential and characterized by a certain
lifetime, that can be calculated from the Fermi golden rule. But if
one shines on the system an intense laser field, tuned at the
transition frequency 3-1, the evolution can be different. This
problem was investigated in Ref.\ \cite{MPS}, where it was found
that the lifetime of the initial state depends on the intensity of
the laser field. In the limit of an extremely intense field, the
initial state undergoes a ``continuous observation" and the decay
should be considerably slowed down (quantum Zeno effect). The aim
of this paper is to study this effect in more detail and discuss a
new phenomenon \cite{FPOlomouc}: we shall see that for physically
sensible values of the intensity of the laser, the decay can be
{\em enhanced}, rather than hindered. This can be viewed as an
``inverse" quantum Zeno effect. An important role in this context
will be played by the specific properties of the interaction
Hamiltonian, in particular by the ``form factor" of the
interaction.

Other authors have studied physical effects that are related to
those we shall discuss. The features of the matrix elements of the
interaction Hamiltonian were investigated in the context of the
quantum Zeno effect by Kofman and Kurizki \cite{Kurizki}, who also
emphasized that different quantum Zeno regimes are present.
Plenio, Knight and Thompson discussed the quantum Zeno effect due
to ``continuous" measurements and considered several physical
systems whose evolution is modified by an external field
\cite{PKT}. There is also work by Kraus on a similar subject
\cite{Kraus}. Finally, Zhu, Narducci and Scully \cite{ZNS}
investigated the electromagnetic-induced transparency in a context
similar to that considered in this paper. In some sense, our
present investigation ``blends" these studies, by taking into
account the important role played by the matrix elements of the
interaction. This will enable us to discuss some new features of
the evolution that have not been considered before. We shall look
at this phenomenon from several perspectives, by first solving the
time-dependent Schr\"odinger equation, then looking at the
spectrum of the emitted photons and finally constructing the
dressed (Fano) states.

Our analysis will be performed within the Weisskopf-Wigner
approximation and no deviations at short
\cite{shortt,Raizen,Hillery} and long \cite{longtt} times will be
considered. The features of the quantum mechanical evolution are
summarized in \cite{temprevi} and have already been discussed
within a quantum field theoretical framework
\cite{BMT,FP1,Jap,Spagna}, where several subtle effects have to be
properly taken into account.

This paper is organized as follows: in Section 2 we introduce the
3-level system bathed in the laser field. Its temporal evolution
is studied in Section 3. The spectrum of the photons emitted
during the evolution is evaluated in Section 4. Section 5 contains
a discussion in terms of dressed states, Section 6 an analysis of
the influence of additional levels on the lifetime and Section 7
some concluding remarks.

\setcounter{equation}{0}
\section{Preliminaries and definitions }
 \label{sec-preldef}
 \andy{preldef}

\begin{figure}
\centerline{\epsfig{file=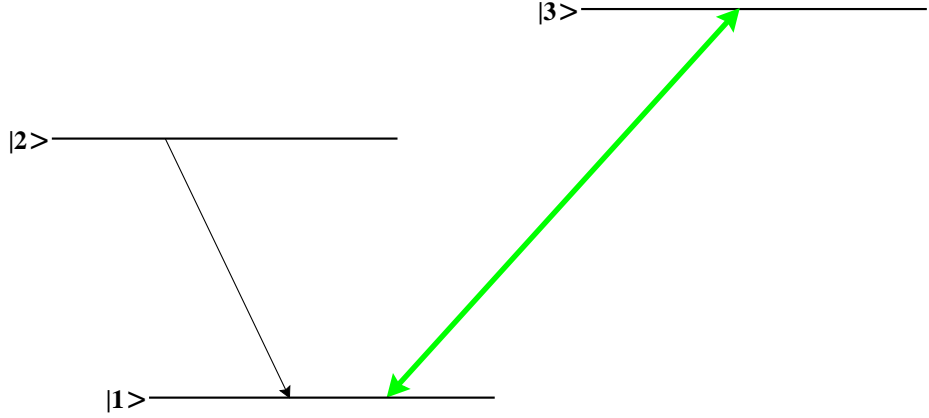,height=5.5cm}}
\caption{Level configuration}
\label{fig:fig0}
\end{figure}

We consider the Hamiltonian ($\hbar=c=1$)\cite{MPS}:
\andy{ondarothamdip6}
\barr
H &=& H_{0}+H_{\rm int}\nonumber\\
&=& \omega_0|2\rangle\langle 2|+\Omega_0|3\rangle\langle 3|
+\sum_{\bmsub k,\lambda}\omega_k a^\dagger_{\bmsub{k}\lambda}
a_{\bmsub{k}\lambda}
+\sum_{\bmsub k,\lambda}\left(\phi_{\bmsub k\lambda}
a_{\bmsub k\lambda}^\dagger|1\rangle\langle2|
+\phi_{\bmsub k\lambda}^* a_{\bmsub
k\lambda}|2\rangle\langle1|\right)\nonumber\\ & &+\sum_{\bmsub
k,\lambda}\left(\Phi_{\bmsub k\lambda}
a_{\bmsub k\lambda}^\dagger|1\rangle\langle3|
+\Phi_{\bmsub k\lambda}^* a_{\bmsub
k\lambda}|3\rangle\langle1|\right),
\label{eq:ondarothamdip6}
\earr
where the first two terms are the free Hamiltonian of the 3-level
atom (whose states $|i\rangle$ $(i=1,2,3)$ have energies $E_1=0$,
$\omega_0=E_2-E_1>0$, $\Omega_0=E_3-E_1>0$), the third term is the
free Hamiltonian of the EM field and the last two terms describe
the $1\leftrightarrow2$ and $1\leftrightarrow3$ transitions in the
rotating wave approximation, respectively. (See Figure
\ref{fig:fig0}.)
States $|2\rangle$ and $|3\rangle$ are chosen so that no transition
between them is possible (e.g., because of selection rules). The
matrix elements of the interaction Hamiltonian read
\andy{intel}
\beq \label{eq:intel}
\phi_{\bmsub k\lambda}=\frac{e}{\sqrt{2\epsilon_0 V\omega}}
\int d^3x\;e^{-i\bmsub k\cdot\bmsub x}\bm\epsilon_{\bmsub k\lambda}^*\cdot \bm
j_{12}(\bm x)\qquad
\Phi_{\bmsub k\lambda}=\frac{e}{\sqrt{2\epsilon_0 V\omega}}
\int d^3x\;e^{-i\bmsub k\cdot\bmsub x}\bm\epsilon_{\bmsub k\lambda}^*\cdot \bm
j_{13}(\bm x),
\eeq
where $-e$ is the electron charge, $\epsilon_0$ the vacuum
permittivity, $V$ the volume of the box, $\omega=|\bm k|$,
$\bm\epsilon_{\bmsub k\lambda}$ the photon polarization and $\bm
j_{\rm fi}$ the transition current of the radiating system. For
example, in the case of an electron in an external field, we have
$\bm j_{\rm fi}=\psi_{\rm f}^\dagger \bm\alpha\psi_{\rm i}$ where
$\psi_{\rm i}$ and $\psi_{\rm f}$ are the wavefunctions of the
initial and final state, respectively, and $\bm\alpha$ is the
vector of Dirac matrices. For the sake of generality we are using
relativistic matrix elements, but our analysis can also be
performed with nonrelativistic ones $\bm j_{\rm fi}=\psi_{\rm
f}^*\bm p\psi_{\rm i}/m_e$, where $\bm p/m_e$ is the electron
velocity.

We shall concentrate our attention on a 3-level system bathed in a
continuous laser beam, whose photons have momentum $\bm k_0$ ($|\bm
k_0|=\Omega_0$) and polarization $\lambda_0$, and assume,
throughout this paper, that
\andy{noint}
\beq
\phi_{\bmsub k_0\lambda_0}=0,
\label{eq:noint}
\eeq
i.e., the laser does not interact with state $|2\rangle$. Let the
laser be in a coherent state $|\alpha_0\rangle$ with a very large
average number $\bar{N}_0=|\alpha_0|^2$ of $\bm k_0$-photons in
volume $V$ [we will eventually consider the thermodynamical limit;
see Eq.~(\ref{eq:limterm})]. In the picture defined by the unitary
operator
\andy{unitaryT}
\beq
T(t)=\exp\left(\alpha_0^* e^{i\Omega_0 t}
a_{\bmsub{k}_0\lambda_0}-\alpha_0 e^{-i\Omega_0 t}
a^\dagger_{\bmsub{k}_0\lambda_0}\right),
\label{eq:unitaryT}
\eeq
the Hamiltonian (\ref{eq:ondarothamdip6}) reads
\andy{newHam}
\beq
H(t)=THT^\dagger+i\dot T T^\dagger=H+\left(\Phi_{\bmsub
k_0\lambda_0}
\alpha_0^* e^{i\Omega_0 t} |1\rangle\langle3|
+\Phi_{\bmsub k_0\lambda_0}^*
\alpha_0 e^{-i\Omega_0 t}|3\rangle\langle1|\right).
\label{eq:newHam}
\eeq
In this picture, the $\bm k_0$ mode is initially in the vacuum
state \cite{Cohen} and by noting that for $\bar N_0\gg 1$
\andy{appros1}
\beq
\left|\langle1; 0_{\bmsub k\lambda}|H(t)
|3; 0_{\bmsub k\lambda}\rangle\right|=\sqrt{\bar N_0}
|\Phi_{\bmsub k_0\lambda_0}|\gg
\left|\langle1;1_{\bmsub k\lambda}|H(t)
|3;0_{\bmsub k\lambda}\rangle\right|=
|\Phi_{\bmsub k\lambda}|, \nonumber \\
\label{eq:appros1}
\eeq
the Hamiltonian (\ref{eq:newHam}) becomes
\andy{newondarothamdip7}
\barr
H & \simeq& \omega_0|2\rangle\langle 2|+\Omega_0|3\rangle\langle 3|
+\sum_{\bmsub k,\lambda}\omega_k {a}^\dagger_
{\bmsub{k}\lambda} {a}_ {\bmsub{k}\lambda} +{\sum_{\bmsub
k,\lambda}}'\left(\phi_{\bmsub k\lambda}
a_{\bmsub k\lambda}^\dagger|1\rangle\langle2|
+\phi_{\bmsub k\lambda}^* a_{\bmsub
k\lambda}|2\rangle\langle1|\right)\nonumber\\ &
&+\left(\Phi_{\bmsub k_0\lambda_0}
\alpha_0^* e^{i\Omega_0 t} |1\rangle\langle3|
+\Phi_{\bmsub k_0\lambda_0}^*
\alpha_0 e^{-i\Omega_0 t}|3\rangle\langle1|\right),
\label{eq:newondarothamdip7}
\earr
where a prime means that the summation does not include $(\bm
k_0,\lambda_0)$ [due to hypothesis (\ref{eq:noint})]. In the above
equations and henceforth, the vector $|i;n_{\bmsub k
\lambda}\rangle$ represents a state in which the atom is in state
$|i\rangle$ and the electromagnetic field in a state with $n_{\bmsub
k \lambda}$ $(\bm k,\lambda)$-photons. We shall analyze the behavior
of the system under the action of a continuous laser beam of high
intensity. Under these conditions, level configurations similar to
that of Figure \ref{fig:fig0} give rise to the phenomenon of induced
transparency \cite{traspindotta}, for laser beams of sufficiently
high intensities. Our interest, however, will be focused on {\em
unstable} initial states: we shall study the temporal behavior of
level \#2 when the system is shined by a continuous laser of
intensity comparable to those used to obtain induced transparency.

Notice that in Eq.\ (\ref{eq:newondarothamdip7}) the spontaneous
decay $3\to 1$  has been neglected with respect to the stimulated
transition, because of the large factor $\sqrt{\bar N_0}\gg 1$ in
Eq. (\ref{eq:appros1}). However, since our interest is primarily
in the first step of this process, namely the decay $2\to1$, these
smaller, later effects (of the order of $1/N_0$) do not change our
conclusions.

The operator
\andy{operatoreN}
\beq
{\cal N}=|2\rangle\langle 2|+{\sum_{\bmsub k,\lambda}}'
{a}^\dagger_{\bmsub{k}\lambda}{a}_{\bmsub{k}\lambda},
\label{eq:operatoreN}
\eeq
satisfies
\beq
[H,{\cal N}]=0,
\eeq
which implies the conservation of the total number of photons plus
the atomic excitation (Tamm-Dancoff approximation
\cite{TammDancoff}). The Hilbert space splits therefore into
sectors that are invariant under the action of the Hamiltonian: in
our case, the system evolves in the subspace labelled by the
eigenvalue ${\cal N}=1$ and the analysis can be restricted to this
sector \cite{Knight}.

\setcounter{equation}{0}
\section{Temporal evolution}
\label{tempevol}
\andy{tempevol}

We will study the temporal evolution by solving the time-dependent
Schr\"odinger equation
\andy{Schrt}
\beq\label{eq:Schrt}
i\frac{d}{dt}|\psi(t)\rangle= H(t)|\psi(t)\rangle,
\eeq
where the states of the total system in the sector ${\cal N}=1$
read
\andy{statesdefin}
\beq\label{eq:statesdefin}
|\psi(t)\rangle=x(t)|2;0\rangle+{\sum_{\bmsub k,\lambda}}'y_{\bmsub
k\lambda}(t)|1;1_{\bmsub k\lambda}\rangle+{\sum_{\bmsub
k,\lambda}}'z_{\bmsub k\lambda}(t) e^{-i\Omega_0 t} |3;1_{\bmsub
k\lambda}\rangle
\eeq
and are normalized:
\andy{normpsi6}
\beq\label{eq:normpsi6}
\langle\psi(t)|\psi(t)\rangle=|x(t)|^2+{\sum_{\bmsub k,\lambda}}'|y_{\bmsub k,
\lambda}(t)|^2+{\sum_{\bmsub k,\lambda}}'|z_{\bmsub k,
\lambda}(t)|^2=1. \qquad (\forall t)
\eeq
By inserting (\ref{eq:statesdefin}) in (\ref{eq:Schrt}) one obtains
the equations of motion
\andy{eqmotoxy6}
\barr
i\,\dot x(t)&=&\omega_0x(t)+{\sum_{\bmsub k,\lambda}}'\phi_{\bmsub
k\lambda}^*y_{\bmsub k\lambda}(t),\nonumber\\ i\,\dot y_{\bmsub
k\lambda}(t)&=&\phi_{\bmsub k\lambda}x(t)+\omega_k y_{\bmsub
k\lambda}(t)+\alpha_0^*\Phi_{\bmsub k_0\lambda_0}z_{\bmsub
k\lambda}(t),\nonumber\\ i\,\dot z_{\bmsub
k\lambda}(t)&=&\alpha_0\Phi^*_{\bmsub k_0\lambda_0}y_{\bmsub
k\lambda}(t)+\omega_k z_{\bmsub k\lambda}(t),
\label{eq:eqmotoxy6}
\earr
where a dot denotes time derivative.  At time $t=0$ we prepare our
system in the state
\andy{condinxy6}
\beq\label{eq:condinxy6}
|\psi(0)\rangle=|2;0\rangle\quad\Leftrightarrow\quad
x(0)=1,\;y_{\bmsub k\lambda}(0)=0,\;z_{\bmsub k\lambda}(0)=0,
\eeq
which is an eigenstate of the free Hamiltonian
\andy{hhoo}
\beq\label{eq:hhoo}
H_0|\psi(0)\rangle=H_0|2;0\rangle=\omega_0|2;0\rangle.
\eeq
Incidentally, we stress that the choice of the initial state is
different from that of Ref.\ \cite{Cook}, where the 3-level atom is
initially in the ground state (\#1) and a Rabi oscillation to level
\#2, provoked by an rf-field, is inhibited by a pulsed laser,
resonating between levels \#1 and \#3, that performs the
``observation" of level \#1. In our case, the atom is initially in
level \#2, so that it can {\em spontaneously} decay to level \#1, and
it is ``continuously observed" by a continuous laser at the 1-3
frequency \cite{MPS}: As soon as the system has decayed to level \#1,
the (intense) laser provokes the $1\to3$ transition. (The
irreversibility inherent in the act of observation is eventually
brought in by the spontaneous decay of level \#3.) This brings us
conceptually closer to the seminal formulation
\cite{shortt,QZE} of quantum Zeno effect.

By Laplace transforming the system of differential equations
(\ref{eq:eqmotoxy6}) and incorporating the initial condition
(\ref{eq:condinxy6}) we get the algebraic system
\andy{eqmototrasf6}
\barr
is\,\wtilde x(s)&=&\omega_0\wtilde x(s)+{\sum_{\bmsub
k,\lambda}}'\phi_{\bmsub k\lambda}^*\wtilde y_{\bmsub
k\lambda}(s)+i,\nonumber\\ is\,\wtilde y_{\bmsub
k\lambda}(s)&=&\phi_{\bmsub k\lambda}\wtilde x(s)+\omega_k\wtilde
y_{\bmsub k\lambda}(s)+\alpha_0^*\Phi_{\bmsub k_0\lambda_0}\wtilde
z_{\bmsub k\lambda}(s),\nonumber\\ is\,\wtilde z_{\bmsub
k\lambda}(s)&=&\alpha_0\Phi^*_{\bmsub k_0\lambda_0}\wtilde
y_{\bmsub k\lambda}(s)+\omega_k \wtilde z_{\bmsub k\lambda}(s),
\label{eq:eqmototrasf6}
\earr
where
\beq
\wtilde f(s)=\int_0^\infty dt\; e^{-st} f(t),\qquad
f(t)=\frac{1}{2\pi i}\int_{\rm B} ds\;e^{ts}\wtilde f(s),
\eeq
the Bromwich path B being a vertical line $\mbox{Re} s=$constant in
the half plane of convergence of the Laplace transform. (Very
similar equations of motion can be obtained by assuming that the
external (laser) field is initially in a number state $N_0$, with
$N_0$ very large \cite{FPOlomouc}. See also the discussion in
Section \ref{drst}.) It is straightforward to obtain
\andy{xs,ys,zs}
\barr
\wtilde x(s)&=&\frac{1}{s+i\omega_0+Q(B,s)},\label{eq:xs}\\
\wtilde y_{\bmsub k\lambda}(s)&=&\frac{-i\phi_{\bmsub k\lambda}(s+i\omega_k)}
{(s+i\omega_k)^2+B^2}\;\wtilde x(s),\label{eq:ys}\\
\wtilde z_{\bmsub k\lambda}(s)&=&-\frac{\sqrt{\bar N_0}\Phi^*_{\bmsub k_0\lambda_0}
\phi_{\bmsub k\lambda}}
{(s+i\omega_k)^2+B^2}\;\wtilde x(s),
\label{eq:zs}
\earr
with
\andy{Q(B,s)dipdiscr}
\beq\label{eq:Q(B,s)dipdiscr}
Q(B,s)=\sum_{\bmsub k,\lambda}|\phi_{\bmsub
k\lambda}|^2\frac{s+i\omega_k}{(s+i\omega_k)^2+B^2}
\eeq
and where
\andy{Bdef}
\beq\label{eq:Bdef}
B^2=\bar N_0\,|\Phi_{\bmsub k_0\lambda_0}|^2
\eeq
is proportional to the intensity of the laser field and can be viewed
as the ``strength" of the observation performed by the laser beam on
level \#2 \cite{MPS}. See the paragraph following Eq.\
(\ref{eq:hhoo}). Note that the coupling $B$ is related to the Rabi
frequency by the simple relation $B=\Omega_{\rm Rabi}/2$.

In the continuum limit ($V\rightarrow\infty$), the matrix elements scale
as follows
\andy{chi}
\beq\label{eq:chi}
\lim_{V\rightarrow\infty}\frac{V\omega^2}{(2\pi)^3}\sum_\lambda\int d \Omega
|\phi_{\bmsub k\lambda}|^2 \equiv g^2\omega_0\chi^2(\omega),
\eeq
where $\Omega$ is the solid angle.
The (dimensionless) function $\chi(\omega)$ and  coupling constant $g$
have the following general properties, discussed
in Appendix A:
\andy{chiprop,g2}
\barr
\chi^2(\omega) & \propto &
\left\{ \begin{array} {ll}
         \omega^{2j\mp 1} &  \quad \mbox{if $\omega \ll \Lambda$}\\
         \omega^{-\beta} & \quad \mbox{if $\omega \gg \Lambda$}
\end{array}
\right. ,
\label{eq:chiprop}\\
g^2 &=& \alpha (\omega_0 /\Lambda )^{2j+1\mp 1} , \label{eq:g2}
\earr
where $j$ is the total angular momentum of the photon emitted in
the $2\rightarrow 1$ transition, $\mp$ represent electric and
magnetic transitions, respectively, $\beta (> 1)$ is a constant,
$\alpha$ the fine structure constant and $\Lambda$ a natural cutoff
(of the order of the inverse size of the emitting system, e.g.\ the
Bohr radius for an atom), which determines the range of the atomic
or molecular form factor \cite{BLP}.

In order to scale the quantity $B$, we take the limit of very large cavity,
by keeping the density of $\Omega_0$-photons in the cavity constant:
\andy{limterm}
\beq\label{eq:limterm}
V\rightarrow\infty,\qquad
\bar N_0\rightarrow\infty,\quad\mbox{with}\quad
\frac{\bar N_0}{V}=n_0=\mbox{const}
\eeq
and obtain from (\ref{eq:Bdef})
\beq
B^2 = n_0 V |\Phi_{\bmsub k_0\lambda_0}|^2 = (2\pi)^3 n_0
|\varphi_{\lambda_0} (\bm k_0)|^2 ,
\eeq
where $\varphi\equiv\Phi V^{1/2}/(2\pi)^{3/2}$ is the scaled matrix
element of the 1-3 transition. As we shall see, in order to affect
significantly the lifetime of level \#2, we shall need a high value
of $B$, namely, a laser beam of high intensity. It is therefore
interesting to consider a 1-3 transition of the dipole type, in which
case the above formula reads
\andy{Bdip}
\beq\label{eq:Bdip}
B^2 = 2\pi\alpha\Omega_0|\bm{\epsilon}_{\bmsub{k_0}\lambda_0}^*
\cdot\bm x_{13}|^2 n_0,
\eeq
where $\bm x_{13}$ is the dipole matrix element.

\subsection{Laser off}
\label{laseroff}
\andy{laseroff}

\begin{figure}
\centerline{\epsfig{file=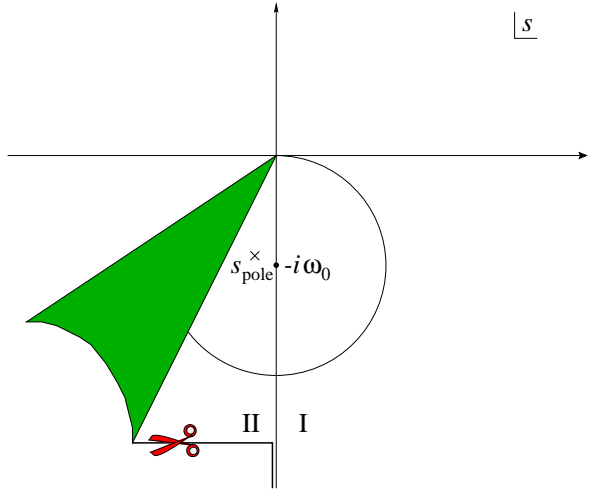,height=6.5cm}}
\caption{%
Cut and pole in the $s$-plane ($B=0$) and convergence circle for
the expansion of $Q(s)$ around $s=-i\omega_0$. I and II are the
first and second Riemann sheets, respectively. The pole is on the
second Riemann sheet, at a distance $O(g^2)$ from $-i\omega_0$.}
\label{fig:fig1}
\end{figure}

Let us first look at the case $B=0$. The laser is off and we expect
to recover the well-known physics of the spontaneous emission a
two-level system prepared in an excited state and coupled to the
vacuum of the radiation field. In this case, $Q(0,s)$ is nothing
but the self-energy function
\andy{sef}
\beq
Q(s)\equiv Q(0,s)=\sum_{\bmsub k,\lambda}|\phi_{\bmsub
k\lambda}|^2\frac{1}{s+i\omega_k},
\label{eq:sef}
\eeq
which becomes, in the continuum limit,
\andy{Q(s)cont}
\beq\label{eq:Q(s)cont}
Q(s) \equiv g^2 \omega_0 q(s)\equiv
-i g^2 \omega_0 \int_0^\infty d\omega
\frac{\chi^2(\omega)}{\omega-is},
\eeq
where $\chi$ is defined in (\ref{eq:chi}). The function $\wtilde
x(s)$ in Eq.\ (\ref{eq:xs}) (with $B=0$) has a logarithmic branch
cut, extending from 0 to $-i\infty$, and no singularities on the
first Riemann sheet (physical sheet) \cite{FP1}. On the other hand,
it has a simple pole on the second Riemann sheet, that is the
solution of the equation
\andy{poleq}
\beq\label{eq:poleq}
s+i\omega_0+g^2\omega_0 q_{\rm II}(s)=0,
\eeq
where
\beq
q_{\rm II}(s)=q(s e^{-2\pi i})=q(s)+2 \pi\chi^2 (is)
\eeq
is the determination of $q(s)$ on the second Riemann sheet. We note
that $g^2 q(s)$ is $O(g^2)$, so that the pole can be found
perturbatively: by expanding $q_{\rm II}(s)$ around $-i\omega_0$ we
get a power series, whose radius of convergence is $R_c=\omega_0$
because of the branch point at the origin. The circle of
convergence lies half on the first Riemann sheet and half on the
second sheet (Figure \ref{fig:fig1}).
The pole is well inside the convergence circle,
because $|s_{\rm pole}+i\omega_0|\sim g^2\omega_0\ll R_c$, and
we can write
\beq
s_{\rm pole}= -i\omega_0-g^2\omega_0 q_{\rm II}(-i\omega_0-0^+)+O(g^4)
=-i\omega_0-g^2 \omega_0 q(-i\omega_0+0^+)+O(g^4),
\eeq
because $q_{\rm II}(s)$ is the analytic continuation of $q(s)$
below the branch cut. By using the formula
\beq
\lim_{\varepsilon\rightarrow 0^+}\frac{1}{x\pm i\varepsilon}=
P\frac{1}{x}\mp i\pi\delta(x),
\eeq
one gets from (\ref{eq:Q(s)cont})
\andy{Q(-ieta)}
\barr
q(-i\eta+0^+)&=&-i\int_0^\infty d\omega\,\chi^2(\omega)\frac{1}
{\omega-\eta-i0^+}\nonumber\\ &=&\pi
\chi^2(\eta)\theta(\eta)-iP\int_0^\infty d\omega\,
\chi^2(\omega)\frac{1} {\omega-\eta}
\label{eq:Q(-ieta)}
\earr
and by setting
\andy{spole}
\beq\label{eq:spole}
s_{\rm pole}=-i\omega_0+i\Delta E-\frac{\gamma}{2},
\eeq
one obtains
\andy{Fgr}
\beq
\gamma=2\pi g^2\omega_0\chi^2(\omega_0)+O(g^4), \qquad \Delta
E=g^2\omega_0 P \int_0^\infty d\omega\;
\frac{\chi^2(\omega)}{\omega-\omega_0} +O(g^4), \label{eq:Fgr}
\eeq
which are the Fermi ``golden rule" and the second order correction
to the energy of level \#2.

The Weisskopf-Wigner approximation \cite{seminal} consists in
neglecting all branch cut contributions and approximating the
self-energy function with a constant (its value in the pole), that is
\andy{WW}
\beq
\wtilde x(s)=\frac{1}{s+i\omega_0+Q(s)}\simeq\frac{1}{s+i\omega_0+Q_{\rm II}(s_{\rm pole})}
=\frac{1}{s-s_{\rm pole}},
\label{eq:WW}
\eeq
where in the last equality we used the pole equation
(\ref{eq:poleq}). This yields a purely exponential behavior,
$x(t)=\exp(s_{\rm pole} t)$, without short-time (and long-time)
corrections. As is well known, the latter are all contained in the
neglected branch cut contribution.

\subsection{Laser on}
\label{laseron}
\andy{laseron}

We turn now our attention to the situation with the laser switched
on ($B\neq0$) and tuned at the 1-3 transition frequency $\Omega_0$.
The self energy function $Q(B,s)$ in (\ref{eq:Q(B,s)dipdiscr})
depends on $B$ and can be written in terms of the self energy
function $Q(s)$ in absence of laser field [Eq.\ (\ref{eq:sef})], by
making use of the following remarkable property:
\andy{propnotev}
\beq
Q(B,s)=\frac{1}{2}\sum_{\bmsub k,\lambda}|\phi_{\bmsub
k\lambda}|^2\left(\frac{1}{s+i\omega_k+iB}+\frac{1}{s+i\omega_k-iB}\right)
= \frac{1}{2}\left[Q(s+iB)+Q(s-iB)\right].
\label{eq:propnotev}
\eeq
Notice, incidentally, that in the continuum limit ($V\to\infty$),
due to the above formula, $Q(B,s)$ scales just like $Q(s)$. The
position of the pole $s_{\rm pole}$ (and as a consequence the
lifetime $\tau_{\rm E}\equiv\gamma^{-1}=-1/2\mbox{Re} s_{\rm
pole}$) depends on the value of $B$. There are now two branch cuts
in the complex $s$ plane, due to the two terms in
(\ref{eq:propnotev}). They lie over the imaginary axis, along
$(-i\infty,-iB]$ and $(-i\infty,+iB]$.

\begin{figure}
\centerline{\epsfig{file=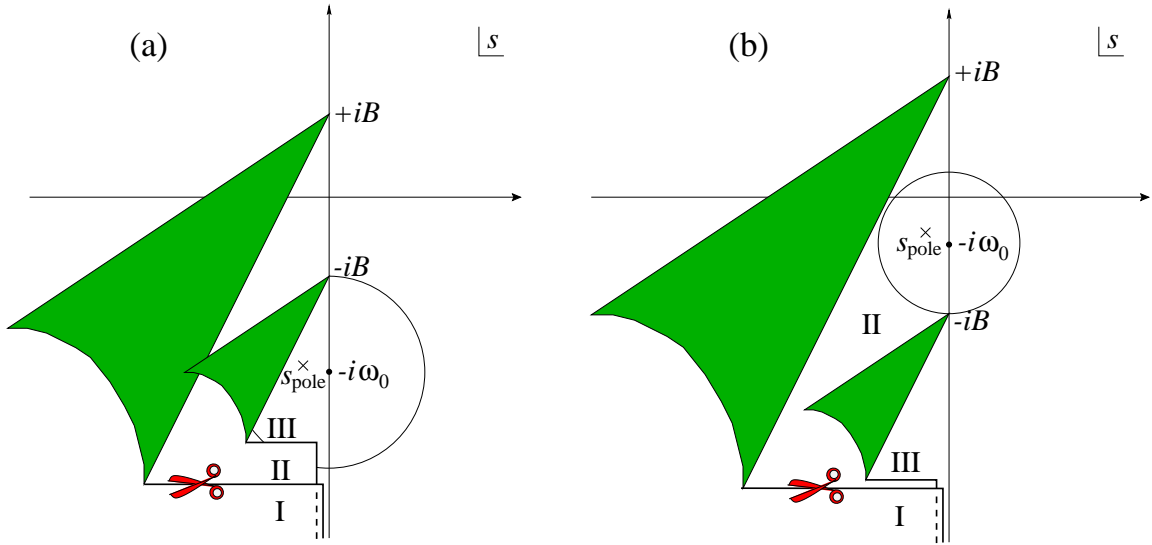,width=\textwidth}}
\caption{%
Cuts and pole in the $s$-plane ($B \neq 0$) and convergence circle
for the expansion of $Q(B,s)$ around $s=-i\omega_0$. I , II and III
are the first, second and third Riemann sheets, respectively. (a)
$B<\omega_0$. (b) $B>\omega_0$. In both cases, the pole is at a
distance $O(g^2)$ from $-i\omega_0$.}
\label{fig:tagli}
\end{figure}

The pole satisfies the equation
\beq
s + i\omega_0 + Q(B, s)=0,
\eeq
where  $Q(B,s)$ is of order $g^2$, as before, and can again be
expanded in power series around $s=-i\omega_0$, in order to find the
pole perturbatively. However, this time one has to choose the right
determination of the function $Q(B,s)$. Two cases are mathematically
possible: a) The branch point $-iB$ is situated above $-i\omega_0$,
so that $-i\omega_0$ lies on both cuts. See Figure
\ref{fig:tagli}(a); b) The branch point $-iB$ is situated below
$-i\omega_0$, so that $-i\omega_0$ lies only on the upper branch cut.
See Figure \ref{fig:tagli}(b). We notice that, although
mathematically conceivable, the latter case ($B>\omega_0$) cannot be
tackled within our approximations, for a number of additional effects
would then have to be considered: multi-photon processes would take
place, the other atomic levels would start to play an important role
and our approach (3-level atom in the rotating wave approximation)
would no longer be valid. We therefore restrict our attention to
values of $B$ that are high (of the same order of magnitude as those
utilized in electromagnetic induced transparency), but not extremely
high, so that our starting approximations still apply.

In case a), i.e.\ for $B<\omega_0$, the pole is on the third Riemann
sheet (under both cuts) and the power series converges in a circle
lying half on the first and half on the third Riemann sheet, within a
convergence radius $R_c=\omega_0-B$, which decreases as $B$ increases
[Figure~\ref{fig:tagli}(a)]. For the sake of completeness we also
notice that in case b), i.e.\ for $B>\omega_0$, the pole would be on
the second Riemann sheet (under the upper cut only) and the power
series would converge in a circle lying half on the first and half on
the second Riemann sheet, within a convergence radius
$R_c=B-\omega_0$, which increases with $B$
[Figure~\ref{fig:tagli}(b)].

In either cases we can write, for $|s_{\rm
pole}+i\omega_0|<R_c=|B-\omega_0|$,
\andy{poloB}
\barr
s_{\rm pole} &=&-i\omega_0-\frac{1}{2}\left\{Q[-i(\omega_0+B)+0^+]
+Q[-i(\omega_0-B)+0^+]\right\}+O(g^4) \nonumber \\
&=&-i \omega_0-\frac{1}{2}g^2 \omega_0\left\{q[-i(\omega_0+B)+0^+]
+q[-i(\omega_0-B)+0^+]\right\}+O(g^4) . \nonumber \\
\label{eq:poloB}
\earr
Equation (\ref{eq:poloB}) enables us to analyze the temporal
behavior of state \#2.

\subsection{Decay rate vs $B$}
\label{decvsB}
\andy{decvsB}

\begin{figure}
\centerline{\epsfig{file=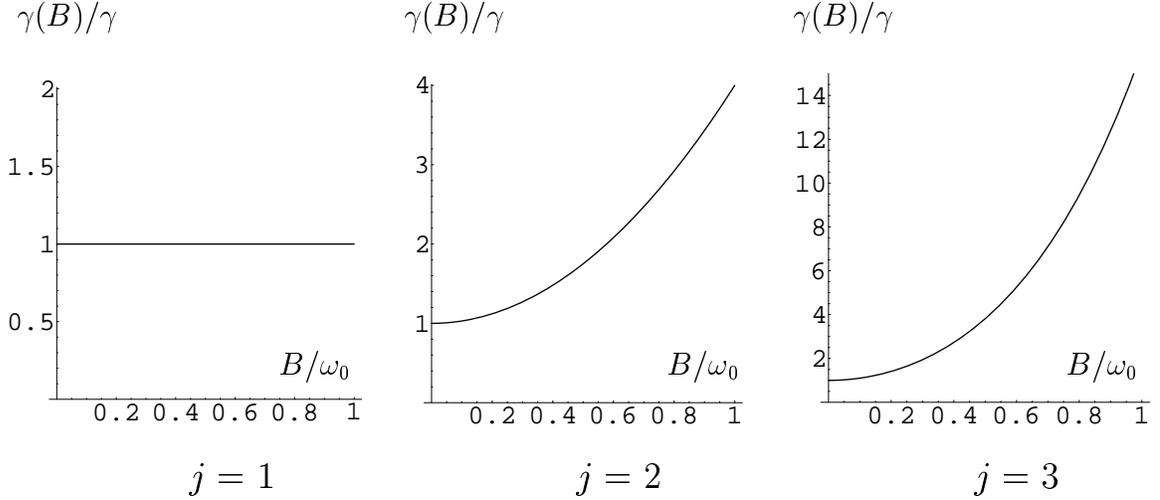,width=\textwidth}}
\caption{%
The decay rate $\gamma(B)$ vs $B$, for electric transitions with
$j=1,2,3$; $\gamma(B)$ is in units $\gamma$ and $B$ in units
$\omega_0$. Notice the different scales on the vertical axis.}
\label{fig:gamma(B)dip}
\end{figure}

We write, as in (\ref{eq:spole}),
\andy{s(B)}
\beq\label{eq:s(B)}
s_{\rm pole}=-i\omega_0+i\Delta E(B)-\frac{\gamma(B)}{2}.
\eeq
Substituting (\ref{eq:Q(-ieta)}) into (\ref{eq:poloB}) and taking
the real part, one obtains the following expression for the decay
rate
\andy{gamma(B)}
\beq\label{eq:gamma(B)}
\gamma(B)=\pi g^2 \omega_0 \left[\chi^2(\omega_0+B)+
\chi^2(\omega_0-B)\theta(\omega_0-B)\right] +O(g^4).
\eeq
On the other hand, by  (\ref{eq:Fgr}), one can write
\andy{MPSs}
\beq\label{eq:MPSs}
\gamma(B)= \gamma \; \frac{\chi^2(\omega_0+B)+
\chi^2(\omega_0-B)\theta(\omega_0-B)}{2\chi^2(\omega_0)} +O(g^4).
\eeq
This is the central result of this paper and involves no
approximations: Equation (\ref{eq:MPSs}) expresses the ``new"
lifetime $\gamma(B)^{-1}$, when the system is bathed in an intense
laser field $B$, in terms of the ``ordinary" lifetime
$\gamma^{-1}$, when there is no laser field. By taking into account
the general behavior (\ref{eq:chiprop}) of the matrix elements
$\chi^2(\omega)$ and substituting into (\ref{eq:MPSs}), one gets to
$O(g^4$)
\andy{gamma(B)dip}
\beq\label{eq:gamma(B)dip}
\gamma(B)\simeq
\frac{\gamma}{2}\left[\left(1+\frac{B}{\omega_0}\right)^{2j\mp 1}
+\left(1-\frac{B}{\omega_0}\right)^{2j\mp 1}
\theta(\omega_0-B)\right], \qquad (B \ll \Lambda)
\eeq
where $\mp$ refers to 1-2 transitions of electric and magnetic type,
respectively. Observe that, since $\Lambda \sim$ inverse Bohr radius,
only the case $B < \omega_0 \ll \Lambda$ is the physically relevant
one
\cite{FPOlomouc}. The decay rate is profoundly modified by the presence of
the laser field. Its behavior is shown in
Figure~\ref{fig:gamma(B)dip} for a few values of $j$. In general, for
$j>1$ (1-2 transitions of electric quadrupole, magnetic dipole or
higher), the decay rate $\gamma(B)$ increases with $B$, so that the
lifetime $\gamma(B)^{-1}$ decreases as $B$ is increased. If one looks
at $B$ as the strength of the ``observation" performed by the laser
beam on level \#2 \cite{MPS}, one can view this phenomenon as an
``inverse" quantum Zeno effect, for decay is {\em enhanced} (rather
than suppressed) by observation.

As we shall see in Sections \ref{tempph} and \ref{drst}, the emitted
photons have different frequencies [for they correspond to decay onto
different dressed (Fano) states]. By selecting the photon wavelength
(i.e., by means of filters), one could therefore also measure the
different contributions to the inverse lifetime in
(\ref{eq:MPSs})-(\ref{eq:gamma(B)dip}). We shall come back to this
point later.

As already emphasized, Eq.\ (\ref{eq:gamma(B)dip}) is valid for
$B\ll\Lambda$. In the opposite (unphysical) case $B\gg\Lambda$, by
(\ref{eq:chiprop}) and (\ref{eq:MPSs}), one gets to O($g^4$)
\andy{gammaom}
\beq\label{eq:gammaom}
\gamma(B) \simeq \frac{\gamma}{2} \; \frac{\chi^2(B)}{\chi^2(\omega_0)}
\propto (B/\Lambda)^{-\beta}.
 \qquad (B \gg \Lambda)
\eeq
This result is similar to that obtained in Ref.\ \cite{MPS}. If such
high values of $B$ were experimentally obtainable, the decay would be
considerably hindered and $B$ could be properly viewed as the
``strength" of the observation performed by the  laser field on level
\#2 (quantum Zeno effect). However, in such a case, many additional
effects would have to be considered and our analysis should be
modified in order to take them into account. A similar remark was
made by Kofman and Kurizki in a different context
\cite{Kurizki}.

A final remark is now in order. If one would use the approximation
(\ref{eq:WW}) in Eq.\ (\ref{eq:propnotev}), in order to evaluate
the new lifetime, i.e. if one sets $Q(s)=Q(s_{\rm pole})=$const,
one would obtain $Q(B,s)=Q(s)=Q(s_{\rm pole})$, i.e.\ no
$B$-dependence. Therefore, the effect we are discussing is
ultimately due to the nonexponential contributions arising from
the cut. In particular, viewed from the perspective of the time
domain, this effect is ascribable to the quadratic short-time
behavior of the $2\to1$ decay.

\subsection{Estimates}
\label{estim}
\andy{estim}

We saw in the previous subsection that the ratio $B/\omega_0$ is the
relevant quantity in the evaluation of the modified lifetime. Let us
therefore try to get a rough feeling for the magnitude of the
relevant physical parameters. In order to affect significantly the
lifetime of level
\#2, we have to look at rather large values of $B$: for instance at 1-3
transition of the electric dipole type. In such a case, Eq.\
(\ref{eq:Bdip}) applies:
\beq
B^2 = 2\pi\alpha\Omega_0|\bm{\epsilon}_{\bmsub{k_0}\lambda_0}^*
\cdot\bm x_{13}|^2 n_0.
\eeq
Considering the angle average
\beq
\langle|\bm{\epsilon}_{\bmsub{k_0}\lambda_0}^*
\cdot\bm x_{13}|^2\rangle=\frac{1}{3}|\bm x_{13}|^2
\eeq
and remembering that the decay rate is
\beq
\Gamma_{13}=\frac{4}{3}\alpha|\bm x_{13}|^2 \Omega_0^3,
\eeq
we obtain
\beq
B^2=\frac{\pi}{2} n_0 \frac{\Gamma_{13}}{\Omega_0^2},
\eeq
which, reinserting $c$'s and $\hbar$'s, reads
\andy{BB2}
\beq\label{eq:BB2}
B^2=\frac{\pi}{2}n_0\hbar\Omega_0\frac{c^3}{\Omega_0^3}\hbar\Gamma_{13}
=(n_0\hbar\Omega_0)\frac{\lambda_L^3}{16\pi^2}(\hbar\Gamma_{13}),
\eeq
where $\lambda_L=2\pi c /\Omega_0$. The quantity $B^2$ has dimensions
of squared energy and is given by the product of the energy of the
laser field contained in the volume $\lambda_L^3/16\pi^2$ times the
energy spread of the $1-3$ transition $\Omega_0$. Therefore $B$
depends on both laser and atomic system. Observe that
$n_0\lambda_L^3$ is the number of laser photons contained in the
volume $\lambda_L^3$.

In terms of laser power $P$ and laser spot area $A$, Eq.\
(\ref{eq:BB2}) reads
\andy{BB1}
\beq\label{eq:BB1}
B^2=\frac{P}{c A}\frac{\lambda_L^3}{16\pi^2}(\hbar\Gamma_{13})
=132 \frac{P\lambda_L^3}{A}(\hbar\Gamma_{13})\;\mbox{eV}^2,
\eeq
where $P$ is expressed in Watt, $\lambda_L$ in $\mu$m, $A$ in
$\mu$m$^2$ and $\hbar\Gamma$ in eV. In Eq.\ (\ref{eq:BB1}) the
quantity $B$ is expressed in suitable units and can be easily
compared to $\omega_0$ [the ratio $B/\omega_0$ being the relevant
quantity in Eq.\ (\ref{eq:gamma(B)dip})]. For laser intensities that
are routinely used in the study of electromagnetic induced
transparency, the effect should be experimentally observable. For a
quick comparison remember that $B$ is just half the Rabi frequency of
the resonant transition $1-3$ [see paragraph following Eq.\
(\ref{eq:Bdef})].

\setcounter{equation}{0}
\section{Photon spectrum}
\label{tempph}
\andy{tempph}

It is interesting to look at the spectrum of the emitted photons.
It is easy to check that, in the Weisskopf-Wigner approximation,
the survival probability
$|\langle\psi(0)|\psi(t)\rangle|^2=|x(t)|^2$ decreases
exponentially with time. The standard way to obtain this result is
to neglect the cut contribution in the complex $s$ plane, or
equivalently, to substitute in (\ref{eq:xs}) the pole determination
of the self-energy function:
\andy{eq:xsexp}
\beq\label{xsexp}
\wtilde x(s)=\frac{1}{s+i\omega_0+Q(B,s)}
\approx\frac{1}{s+i\omega_0+Q(B,s_{\rm pole})} = \frac{1}{s-s_{\rm pole}},
\eeq
from which one gets
\beq
x(t)=\exp(s_{\rm pole}t)=\exp\left(-i\bar\omega_0
t-\frac{\gamma(B)}{2}t\right),
\eeq
where $\bar\omega_0=\omega_0-\Delta E(B)$. In this approximation,
for any value of $B$, the spectrum of the emitted photons is
Lorentzian. The proof is straightforward and is given in Appendix
B. One finds that, for $B=0$, the probability to emit a photon in
the range ($\omega,\omega+ d \omega$) reads
\andy{Lprof}
\beq
dP_{B=0}= g^2 \omega_0
\chi^2(\omega)f_L(\omega-\bar\omega_0;\gamma)d\omega,
\label{eq:Lprof}
\eeq
where
\andy{lorentz1}
\beq\label{eq:lorentz1}
f_L(\omega;\gamma)=\frac{1} {\omega^2+\gamma^2/4}.
\eeq
\begin{figure}
\centerline{\epsfig{file=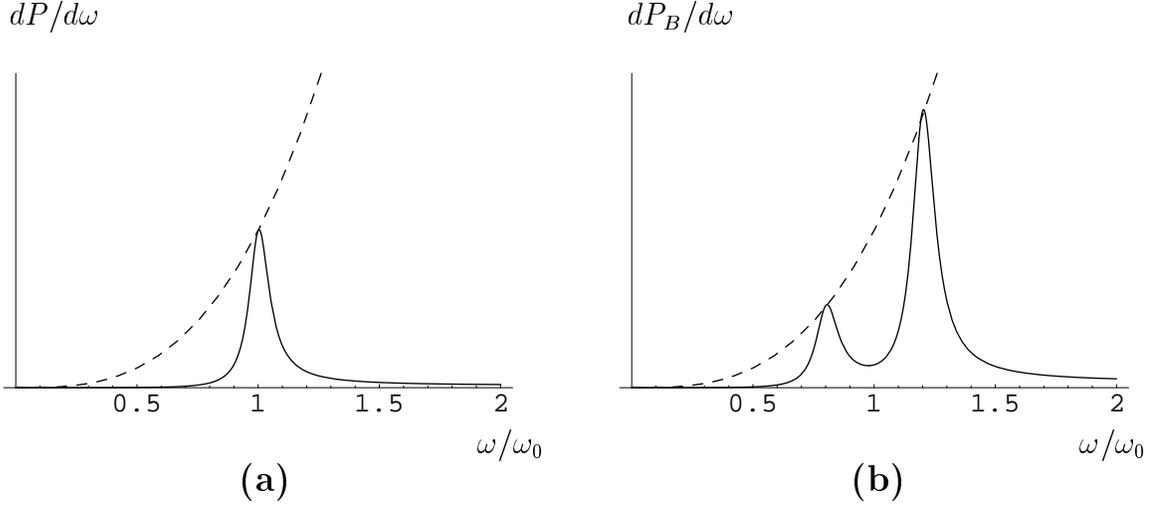,width=\textwidth}}
\caption{The spectrum (\ref{eq:LLprof})
of the emitted photons. The height of the Lorentzians is
proportional to the matrix element $\chi^2(\omega)$ (dashed line).
We chose an electric quadrupole transition, with $j=2$ and $\gamma=
10^{-1}
\bar\omega_0$, and used arbitrary units on the vertical axis. a)
B=0; b) $B=\bar\omega_0/5$; note
that from (\ref{eq:gamma(B)dip}) $\gamma (B)= (28/25)\gamma$.}
\label{fig:Lorentz}
\end{figure}
On the other hand, when $B\neq 0$ one gets:
\andy{LLprof}
\beq
dP_B=g^2 \omega_0 \chi^2(\omega)
\frac{1}{2}
\left[f_L(\omega-\bar\omega_0-B;\gamma(B))+
f_L(\omega-\bar\omega_0+B;\gamma(B))\right]d\omega.
\label{eq:LLprof}
\eeq
The emission probability is given by the sum of two Lorentzians,
centered in $\bar \omega_0 \pm B$. We see that the emission
probability of a photon of frequency $\bar\omega_0+ B$
($\bar\omega_0- B$) increases (decreases) with $B$
(Figure~\ref{fig:Lorentz}). The linewidths are modified according to
Eq.\ (\ref{eq:gamma(B)dip}). When $B$ reaches the ``threshold" value
$\bar\omega_0$, only the photon of higher frequency
$(\bar\omega_0+B)$ is emitted (with increasing probability vs $B$).

Photons of different frequencies are therefore emitted with different
rates. We shall understand better the features of the emission in the
next section, by looking at the dressed states of the system.

\setcounter{equation}{0}
\section{Dressed states and links with induced transparency}
\label{drst}
\andy{drst}

It is useful and interesting to look at our results from a
different perspective, by analyzing the modifications of the energy
levels provoked by the laser field. For simplicity, since the
average number $\bar N_0$ of $\bm k_0$-photons in the total volume
$V$ can be considered very large, we shall perform our analysis in
terms of number (rather than coherent) states of the
electromagnetic field. In this limit,
\andy{appros}
\beq
\langle1; 0_{\bmsub k\lambda}, N_0|H_{\rm int}
|3; 0_{\bmsub k\lambda},N_0-1\rangle=\sqrt{N_0}
\Phi_{\bmsub k_0\lambda_0}\gg
\langle1;1_{\bmsub k\lambda}, N_0-1|H_{\rm int}
|3;0_{\bmsub k\lambda},N_0-1\rangle=
\Phi_{\bmsub k\lambda}, \nonumber \\
\label{eq:appros}
\eeq
$\forall(\bm k,\lambda)\neq(\bm k_0,\lambda_0)$. [This is
equivalent to (\ref{eq:appros1}).] In the above equation and
henceforth, the vector $|i;n_{\bmsub k \lambda}, M_0\rangle$
represents an atom in state $|i\rangle$, with $n_{\bmsub k
\lambda}$ $(\bm k,\lambda)$-photons and $M_0$ laser photons.

In the above approximation, the Hamiltonian
(\ref{eq:ondarothamdip6}) becomes
\andy{ondarothamdip7}
\barr
H & \simeq& \omega_0|2\rangle\langle 2|+\Omega_0|3\rangle\langle 3|
+\sum_{\bmsub k,\lambda}\omega_k {a}^\dagger_
{\bmsub{k}\lambda} {a}_ {\bmsub{k}\lambda} +{\sum_{\bmsub
k,\lambda}}'\left(\phi_{\bmsub k\lambda}
a_{\bmsub k\lambda}^\dagger|1\rangle\langle2|
+\phi_{\bmsub k\lambda}^* a_{\bmsub
k\lambda}|2\rangle\langle1|\right)\nonumber\\ &
&+\left(\Phi_{\bmsub k_0\lambda_0}
a_{\bmsub k_0\lambda_0}^\dagger|1\rangle\langle3|
+\Phi_{\bmsub k_0\lambda_0}^* a_{\bmsub
k_0\lambda_0}|3\rangle\langle1|\right),
\label{eq:ondarothamdip7}
\earr
where a prime means that the summation does not include $(\bm
k_0,\lambda_0)$ [due to hypothesis (\ref{eq:noint})]. Besides
(\ref{eq:operatoreN}), there is now another conserved quantity:
indeed the operator
\andy{operatoreN0}
\beq
{\cal N}_0=|3\rangle\langle 3|+{a}^\dagger_{\bmsub{k}_0\lambda_0}
{a}_{\bmsub{k}_0\lambda_0}
\label{eq:operatoreN0}
\eeq
satisfies
\beq
[H,{\cal N}_0]=[{\cal N}_0,{\cal N}]=0.
\eeq
In this case, the system evolves in the subspace labelled by the
two eigenvalues ${\cal N}=1$ and ${\cal N}_0=N_0$, whose states
read
\label{statesdefin1}
\beq\label{eq:statesdefin1}
|\psi(t)\rangle=x(t)|2;0,N_0\rangle+{\sum_{\bmsub
k,\lambda}}'y_{\bmsub k\lambda}(t)|1;1_{\bmsub
k\lambda},N_0\rangle+{\sum_{\bmsub k,\lambda}}'z_{\bmsub
k\lambda}(t)|3;1_{\bmsub k\lambda},N_0-1\rangle .
\eeq
By using the Hamiltonian (\ref{eq:ondarothamdip7}) and the states
(\ref{eq:statesdefin1}) and identifying $N_0$ with $\bar
N_0=|\alpha_0|^2$, the Schr\"odinger equation yields again the
equations of motion (\ref{eq:eqmotoxy6}), obtained by assuming a
coherent state for the laser mode. Our analysis is therefore
independent of the statistics of the driving field, provided it is
sufficiently intense, and the (convenient) use of number states is
completely justified.

Energy conservation implies that if there are two emitted photons
with different energies (as we saw in the previous section), there
are two levels of different energies to which the atom can decay.
This can be seen by considering the laser-dressed (Fano) atomic
states \cite{Fano}. The shift of the dressed states can be obtained
directly from the structure of the Hamiltonian
(\ref{eq:ondarothamdip7}). In the sector ${\cal N}_0=N_0$, the
operator ${\cal N}_0$ is proportional to the unit operator, the
constant of proportionality being its eigenvalue. Hence one can write
the Hamiltonian in the following form
\andy{okey}
\beq
H=H-\Omega_0{\cal N}_0+\Omega_0N_0,
\label{eq:okey}
\eeq
which, by the setting $E_1+N_0\Omega_0=0$, reads
\andy{tothamN0}
\barr
H &=& H_0+H_{\rm int}\nonumber\\
&=& \omega_0|2\rangle\langle 2| +{\sum_{\bmsub
k,\lambda}}'\omega_k a^\dagger_{\bmsub{k}\lambda} a_
{\bmsub{k}\lambda} +{\sum_{\bmsub k,\lambda}}'\left(\phi_{\bmsub
k\lambda}
a_{\bmsub k\lambda}^\dagger|1\rangle\langle2|
+\phi_{\bmsub k\lambda}^* a_{\bmsub
k\lambda}|2\rangle\langle1|\right)\nonumber\\ &
&+\left(\Phi_{\bmsub k_0\lambda_0}
a_{\bmsub k_0\lambda_0}^\dagger|1\rangle\langle3|
+\Phi_{\bmsub k_0\lambda_0}^* a_{\bmsub
k_0\lambda_0}|3\rangle\langle1|\right).
\label{eq:tothamN0}
\earr
On the other hand, in the sector ${\cal H}_{{\cal N}{\cal N}_0}$
with ${\cal N}=1$ and ${\cal N}_0=N_0$, the last term becomes
\beq
\left(\Phi_{\bmsub k_0\lambda_0}
a_{\bmsub k_0\lambda_0}^\dagger|1\rangle\langle3|
+\Phi_{\bmsub k_0\lambda_0}^* a_{\bmsub
k_0\lambda_0}|3\rangle\langle1|\right)=
\left(\Phi_{\bmsub k_0\lambda_0}
\sqrt{N_0}|1\rangle\langle3|
+\Phi_{\bmsub k_0\lambda_0}^*\sqrt{N_0}|3\rangle\langle1|\right).
\eeq
Let us diagonalize this operator, i.e.\ let us look for two
non-interacting states $|+\rangle$ and $|-\rangle$ which are linear
combinations of the old states $|1\rangle$ and $|3\rangle$. To this
end we write
\andy{rotaz}
\barr
|1\rangle&=&\frac{1}{\sqrt{2}}\left(|+\rangle+e^{i\delta}|-\rangle\right),
\nonumber\\
|3\rangle&=&\frac{e^{i\alpha}}{\sqrt{2}}\left(|+\rangle-
e^{i\delta}|-\rangle\right),
\label{eq:rotaz}
\earr
with $|+\rangle$ e $|-\rangle$ orthonormal:
\beq
\langle +|+\rangle=\langle -|-\rangle=1 ,\qquad \langle
+|-\rangle=0.
\eeq
Plugging (\ref{eq:rotaz}) into (\ref{eq:tothamN0}), the interaction
term becomes
\barr
H_{\rm int}&=&{\sum_{\bmsub
k,\lambda}}'\left[\left(\frac{\phi_{\bmsub k\lambda}}{\sqrt{2}}
a_{\bmsub k\lambda}^\dagger|+\rangle\langle2| +\frac{\phi_{\bmsub
k\lambda}^*}{\sqrt{2}} a_{\bmsub k\lambda}|2\rangle\langle
+|\right) \right.
\nonumber\\ & & \left.
+ \left(\frac{\phi_{\bmsub k\lambda}}{\sqrt{2}}e^{i\delta}
a_{\bmsub k\lambda}^\dagger|-\rangle\langle2| +\frac{\phi_{\bmsub
k\lambda}^*}{\sqrt{2}}e^{-i\delta} a_{\bmsub
k\lambda}|2\rangle\langle
-|\right)\right]\nonumber\\ & &
+B\left[\frac{e^{i\beta}}{2}e^{-i\alpha}
\left(|+\rangle+e^{i\delta}|-\rangle\right)
\left(\langle +|-e^{-i\delta}\langle -|\right)+
\mbox{h.c.}\right],
\earr
where we have set $\Phi_{\bmsub k_0\lambda_0}\sqrt{N_0}=B
e^{i\beta}$. Rearranging the last term
\barr
& &B\bigg[\frac{e^{i(\beta-\alpha)}}{2}
\left(|+\rangle\langle +|-|-\rangle\langle -|+e^{i\delta}
|-\rangle\langle +|-e^{-i\delta}|+\rangle\langle
-|\right)\nonumber\\& &\;\;+\frac{e^{-i(\beta-\alpha)}}{2}
\left(|+\rangle\langle +|-|-\rangle\langle -|+e^{-i\delta}
|+\rangle\langle -|
-e^{i\delta} |-\rangle\langle +| \right)\bigg]
\nonumber\\
&=& B\cos(\beta-\alpha)\left(|+\rangle\langle +|-|-\rangle\langle
-|\right)+iB\sin(\beta-\alpha)\left(e^{i\delta}|-\rangle\langle +|
-e^{-i\delta} |+\rangle\langle -|\right)\nonumber\\
\earr
and setting $\alpha=\beta$ the two states $|+\rangle$ and
$|-\rangle$ decouple and one gets
\barr
H_{\rm int}&=&{\sum_{\bmsub
k,\lambda}}'\left[\left(\frac{\phi_{\bmsub k\lambda}}{\sqrt{2}}
a_{\bmsub k\lambda}^\dagger|+\rangle\langle2| +\frac{\phi_{\bmsub
k\lambda}^*}{\sqrt{2}} a_{\bmsub k\lambda}|2\rangle\langle
+|\right) \right.
\nonumber\\ & & \left.
+ \left(\frac{\phi_{\bmsub k\lambda}}{\sqrt{2}}e^{i\delta}
a_{\bmsub k\lambda}^\dagger|-\rangle\langle2| +\frac{\phi_{\bmsub
k\lambda}^*}{\sqrt{2}}e^{-i\delta} a_{\bmsub
k\lambda}|2\rangle\langle
-|\right)\right]\nonumber\\ & & +B|+\rangle\langle
+|-B|-\rangle\langle -|.
\earr
Therefore we can write
\beq
 H_0+ H_{\rm int}= H'_0+ H'_{\rm int},
\eeq
where the primed free and interaction Hamiltonians read
respectively
\andy{H'0,H'int}
\barr
\label{eq:H'0}
 H'_0&=&\omega_0|2\rangle\langle 2|+B|+\rangle\langle +|
-B|-\rangle\langle -|+{\sum_{\bmsub
k,\lambda}}'\omega_k {a}^\dagger_ {\bmsub{k}\lambda}{a}_
{\bmsub{k}\lambda}, \nonumber \\
 H'_{\rm int}&=&{\sum_{\bmsub k,\lambda}}'\left[\left(\frac{\phi_{\bmsub
k\lambda}}{\sqrt{2}}
 a_{\bmsub k\lambda}^\dagger|+\rangle\langle2|
+\frac{\phi_{\bmsub k\lambda}^*}{\sqrt{2}} a_{\bmsub
k\lambda}|2\rangle\langle +|\right)+
\left(\frac{\phi_{\bmsub
k\lambda}}{\sqrt{2}}
 a_{\bmsub k\lambda}^\dagger|-\rangle\langle2|
+\frac{\phi_{\bmsub k\lambda}^*}{\sqrt{2}} a_{\bmsub
k\lambda}|2\rangle\langle -|\right)\right]\nonumber\\
\label{eq:H'int}
\earr
and we set $\delta=0$. We see that the laser dresses the states
$|1\rangle$ and $|3\rangle$, which (if one includes the $\Omega_0$
photon) are degenerate [with energy $E=0$, due to the choice of
the zero of energy: see line after (\ref{eq:okey})]. The dressed
states $|+\rangle$ and $|-\rangle$ have energies $+B$ and $-B$ and
interact with state $|2\rangle$ with a coupling $\phi_{\bmsub
k\lambda}/\sqrt{2}$. Since $2B=\Omega_{\rm Rabi}$ these are the
well-known Autler-Townes doublet \cite{Autler}.

\begin{figure}
\centerline{\epsfig{file=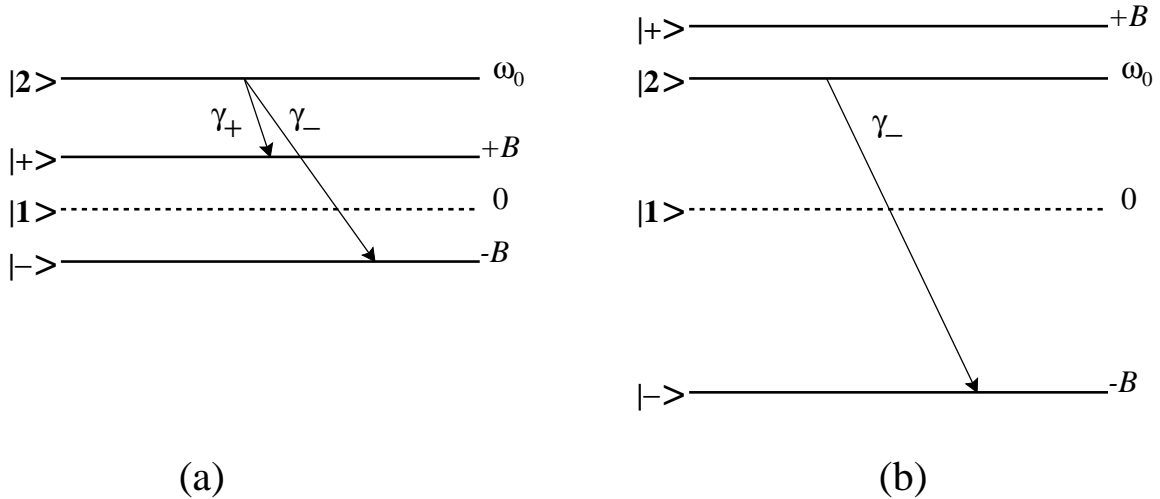,width=\textwidth}}
\caption{Shift of the dressed states $|+\rangle$ and
$|-\rangle$ vs $B$. (a) For $B<\omega_0$ there are two decay
channels, with $\gamma_->\gamma_+$. (b) For $B>\omega_0$ level
$|+\rangle$ is above level $|2\rangle$ and only the
$\gamma_-$ decay channel remains.}
\label{fig:livelli}
\end{figure}

Therefore, by applying the Fermi golden rule, the decay rates into
the dressed states read
\andy{gamaa}
\beq
\gamma_+=2\pi g^2 \omega_0 \frac{\chi^2(\omega_0-B)}{2}\qquad
\gamma_-=2\pi g^2 \omega_0 \frac{\chi^2(\omega_0+B)}{2}
\label{eq:gamaa}
\eeq
and the total decay rate of state $|2\rangle$ is given by their sum
\andy{gambb}
\beq
\gamma=\gamma_++\gamma_-,
\label{eq:gambb}
\eeq
which yields (\ref{eq:gamma(B)}). One sees why there is a threshold
at $B=\omega_0$: For $B<\omega_0$, the energies of both dressed
states $|\pm\rangle$ is lower than that of the initial state
$|2\rangle$ [Figure~\ref{fig:livelli}(a)]. The decay rate $\gamma_-$
increases with $B$, whereas $\gamma_+$ decreases with $B$; their sum
$\gamma$ increases with $B$. These two decays (and their lifetimes)
could be easily distinguished by selecting the frequencies of the
emitted photons, e.g.\ by means of filters.

We also notice, for completeness, that when $B>\omega_0$, the energy
of the dressed state $|+\rangle$ is larger than that of state
$|2\rangle$ and this decay channel disappears
[Figure~\ref{fig:livelli}(b)]. As repeatedly emphasized, this
situation is unphysical and would require a different analysis, for
additional effects would play an important role.

Finally, let us emphasize that if state $|2\rangle$ were {\em below}
state $|1\rangle$, our system would become a three-level system in a
ladder configuration, and the shift of the dressed states would give
rise to electromagnetically induced transparency \cite{traspindotta}.
The situation we consider and the laser power required to bring these
effects to light are therefore similar to those used in induced
transparency.

\setcounter{equation}{0}
\section{Influence of other levels}
\label{many} \andy{many}

Let us now see how our results are modified by the presence of
off-resonant levels. To this end we generalize the three-level
Hamiltonian (\ref{eq:newondarothamdip7}) by including other
off-resonant levels $\ket{j}$ $(j=4,\cdots,N)$ in our analysis:
\andy{manyham}
\barr
H &=& \omega_0|2\rangle\langle
2|+\sum_{j=3}^{N}\Omega_j|j\rangle\langle j| +\sum_{\bmsub
k,\lambda}\omega_k {a}^\dagger_ {\bmsub{k}\lambda} {a}_
{\bmsub{k}\lambda} +{\sum_{\bmsub k,\lambda}}'\left(\phi_{\bmsub
k\lambda} a_{\bmsub k\lambda}^\dagger|1\rangle\langle2|
+\phi_{\bmsub k\lambda}^* a_{\bmsub
k\lambda}|2\rangle\langle1|\right)\nonumber\\
 & &+\sum_{j=3}^{N}\left(\Phi_{j} \alpha_0^* e^{i\Omega_3 t}
|1\rangle\langle j| +\Phi_{j}^* \alpha_0 e^{-i\Omega_3
t}|j\rangle\langle1|\right), \label{eq:manyham}
\earr
where $\Phi_j=\Phi_{j,\bmsub k_0\lambda_0}$ are the matrix
elements of the $1\leftrightarrow j$ transitions and
$\Omega_j=E_j-E_1$ the energy of level $\ket{j}$ [in particular,
$\Phi_{3,\bmsub k_0\lambda_0}=\Phi_{\bmsub k_0\lambda_0}$ and
$\Omega_3=\Omega_0$ in Eq.~(\ref{eq:newondarothamdip7})].

By a calculation similar to that used in Section \ref{tempevol} one
gets again the expression (\ref{eq:xs}) for the Laplace trasform of
the survival amplitude, with the new self-energy function modified by
the presence of other levels
 \andy{manyQ(B,s)}
\beq\label{eq:manyQ(B,s)}
Q(B,s)=\sum_{\bmsub k,\lambda}\frac{|\phi_{\bmsub
k\lambda}|^2}{s+i\omega_k+B^2\sum_{j=3}^{N}\frac{f_j}{s+i\delta_j+i\omega_k}}
,
\eeq
where $f_j=|\Phi_j|^2/|\Phi_3|^2$ and
$\delta_j=\Omega_j-\Omega_3$.

The denominator of $Q(B,s)$ is now a polynomial of order $N-1$ (when
$N=3$ one reobtains Eq.~(\ref{eq:Q(B,s)dipdiscr}) with a quadratic
polynomial). Hence the new $Q(B,s)$ in (\ref{eq:manyQ(B,s)}) has
$N-1$ branching points and the property (\ref{eq:propnotev}) is
generalized to
 \andy{manyprop}
\beq
 Q(B,s)=c_{+}Q(s+i\sigma_+)+c_{-}Q(s+i\sigma_-)
 + \sum_{j=4}^{N} c_j Q(s+i\sigma_j) ,
 \label{eq:manyprop}
\eeq
where $\{-i\sigma_+,-i\sigma_-,-i\sigma_j\}$ $(j=4,\cdots,N)$ are the
branching points, i.e.\ the zeroes of the denominator of $Q(B,s)$. In
this case one has to solve an algebraic equation of $(N-1)$th order,
whose zeroes do not have in general an analytical expression. We seek
a perturbative solution in $B$. It is lengthy, but straightforward,
to obtain up to second order in $B$
 \andy{smallB}
\beq
 \left\{\begin{array}{l}
   \sigma_\pm =\pm B
 -B^2\sum_{j=4}^{N}\frac{f_j}{2\delta_j} \\
 \\
    \sigma_j=\delta_j + B^2 \frac{f_j}{\delta_j} \
 \end{array}\right. ,\qquad
 \left\{\begin{array}{l}
  c_\pm=1/2\mp B\sum_{j=4}^{N} \frac{f_j}{4\delta_j}
 -B^2\sum_{j=4}^{N} \frac{f_j}{2\delta_j^2} \\
 \\
   c_j=B^2 \frac{f_j}{\delta_j^2} \
 \end{array}\right. .
 \label{eq:smallB}
\eeq
{}From the above equations we see that the presence of off-resonant
levels modifies the energies $\sigma_\pm=\pm B$ of the two dressed
states by a shift of order $B^2$ and creates $N-3$ new dressed states
with energies $\delta_j \simeq \Omega_j-\Omega_3$, whose contribution
to the self-energy function is of order $B^2$.

By starting with the self-energy function (\ref{eq:manyprop}) and
looking for the location of the pole one obtains instead of Eq.\
(\ref{eq:gamma(B)dip}) the following expression for the modified
decay rate
 \andy{manygamma}
\barr
 \label{eq:manygamma} \gamma_{\rm many}(B)&=& \gamma
 \left[c_-\left(1-\frac{\sigma_-}{\omega_0}\right)^{\kappa}
 +c_+\left(1-\frac{\sigma_+}{\omega_0}\right)^{\kappa}
 \theta(\omega_0-\sigma_+)\right.\nonumber\\
 & &\qquad\qquad +\left.\sum_{\ell=4}^{N} c_\ell\left(1-\frac{\sigma_\ell}{\omega_0}\right)^{\kappa}
 \theta(\omega_0-\sigma_\ell)\right],
\earr
where $\kappa=2j\mp1$.

By substituting the expressions (\ref{eq:smallB}) for the zeroes and
the coefficients, valid up to second order in $B$, into Eq.\
(\ref{eq:manygamma}) one gets
 \andy{manygamma1}
\barr
 \label{eq:manygamma1}
 \gamma_{\rm many}(B)&\simeq&
\gamma\left\{1+\kappa\frac{B^2}{\omega_0^2}+\frac{B^2}{\omega_0^2}
 \sum_{\ell=4}^{N} f_\ell\left[\left(\kappa\frac{\omega_0}{\delta_\ell}-\frac{\omega_0^2}{\delta_\ell^2}\right)
 -\left(\frac{\omega_0}{\delta_\ell}-\frac{\omega_0^2}{\delta_\ell^2}\right)\theta(\omega_0-\delta_\ell)
\right]\right\}\nonumber\\
 &=& \gamma(B)+\gamma\frac{B^2}{\omega_0^2}
 \sum_{\ell=4}^{N} f_\ell\left[\left(\kappa\frac{\omega_0}{\delta_\ell}-\frac{\omega_0^2}{\delta_\ell^2}\right)
 -\left(\frac{\omega_0}{\delta_\ell}-\frac{\omega_0^2}{\delta_\ell^2}\right)\theta(\omega_0-\delta_\ell)
\right]+{\rm O}(B^3),\nonumber\\
\earr
where $\gamma(B)$ is the decay rate~(\ref{eq:gamma(B)dip}) evaluated
in the three-level approximation.

The above general expression can be evaluated in practical cases of
interest. For instance, by assuming that the off-resonant levels are
well separated from the three main levels, that is by assuming
$\delta_\ell=\Omega_\ell-\Omega_3
>\omega_0$, all dressed states other than $\ket{\pm}$ do not enter in
Eq.~(\ref{eq:manygamma}), because their energies are larger than the
energy $\omega_0$ of level $\ket{2}$, and Eq.~(\ref{eq:manygamma1})
reads
 \andy{manygamma2}
\barr
 \label{eq:manygamma2}
 \gamma_{\rm many}(B)\simeq
\gamma\left[1+\kappa\frac{B^2}{\omega_0^2}\left(1+\sum_{\ell=4}^{N}
f_\ell\frac{\omega_0}{\delta_\ell}\right)\right]
 \simeq\gamma\left(B^*\right),
\earr
where
 \andy{eq:Bstar}
\beq\label{eq:Bstar}
B^*=B\left[1+\sum_{\ell=4}^{N}
f_\ell\frac{\omega_0}{2\delta_\ell}\right].
\eeq
This is the correction sought: the effect of sufficiently
off-resonant levels, $\delta_\ell>\omega_0$, modifies the decay
rate (\ref{eq:gamma(B)dip}), calculated in the three-level
approximation, simply by changing $B$ into $B^*$. Observe that
$f_\ell$ is a rapidly decreasing function of $\ell$ (polynomial
fall-off in atomic systems). Notice also that $B^*>B$, hence the
presence of the other levels enhances the effect discussed in
Section\ \ref{tempevol}.

\setcounter{equation}{0}
\section{Concluding remarks}
\label{concrem} \andy{concrem}

We have studied the evolution of an unstable system under the action
of an external (laser) field. The dynamical evolution of level \#2
(initial state) is modified by the laser field, tuned at the
transition frequency 1-3. For physically sensible values of the
parameters, the decay of level \#2 is {\em faster} when the laser is
present. Equations (\ref{eq:MPSs})-(\ref{eq:gamma(B)dip}) (valid to
4th order in the coupling constant) express the new lifetime as a
function of the ``natural" one and other parameters characterizing
the physical system. The initial state decays to the laser-dressed
states with different lifetimes. We have obtained Eq.\
(\ref{eq:MPSs}) in 3 different ways, deriving the Fermi golden rule
from the time-dependent Schr\"odinger equation, by making use of
Laplace transforms, as in Section
\ref{decvsB}, or starting from the dressed states, as in
(\ref{eq:gamaa})-(\ref{eq:gambb}), or as a consequence of a
normalization condition, as in (\ref{eq:gamcc}). We also computed, in
Section 6, the corrections due to off-resonant levels. We emphasize
that, since we always work in the Weisskopf-Wigner approximation, the
conceptual problems related to state preparation
\cite{FP1} and deviation from exponential behavior
\cite{temprevi,shortt,Hillery,longtt} were not considered.

In which sense is the phenomenon discussed in this paper an
``inverse" quantum Zeno effect? If the situation $B \gg \Lambda$
were experimentally attainable, then decay would be hindered and
one could reasonably speak of a quantum Zeno effect provoked by
the ``continuous" observation performed on the system by the laser
beam. On the other hand, when $B \ll \Lambda$, one can still think
in terms of a ``continuous gaze" of the laser on the system, but
this enhances (rather than hinder) decay. One should also notice
that the inclusion of the spontaneous decay of level \#3 in the
Hamiltonian (\ref{eq:newondarothamdip7}) would not change our
conclusion (up to order $\Gamma_{13}/B$). The interpretation in
terms of an ``inverse" quantum Zeno effect is appealing and
enables one to look at the problem from a different perspective.

\vspace{.5cm}
\noindent {\bf Acknowledgments}
We thank E. Mihokova, G. Scamarcio and L.S.\ Schulman for useful
discussions.


\renewcommand{\thesection}{\Alph{section}}
\setcounter{section}{1}
\setcounter{equation}{0}
\section*{Appendix A}
\label{sec-appA}
\andy{appA}

\renewcommand{\thesection}{\Alph{section}}
\renewcommand{\thesubsection}{{\it\Alph{section}.\arabic{subsection}}}
\renewcommand{\theequation}{\thesection.\arabic{equation}}
\renewcommand{\thefigure}{\thesection.\arabic{figure}}

We discuss here some general properties of the matrix elements and
derive Eqs.~(\ref{eq:chiprop})-(\ref{eq:g2}) of the text. An
exhaustive analysis of some general features of the matrix elements
can be found in \cite{BLP}, but we will focus here on the behavior
at small and large values of $\omega$. The matrix elements
(\ref{eq:intel}) of the 1-2 interaction Hamiltonian read
\andy{intel1}
\beq \label{eq:intel1}
\phi_{\bmsub k\lambda}=\sqrt{\frac{2\pi\alpha}{V\omega}}
\int d^3x\;e^{-i\bmsub k\cdot\bmsub x}\bm\epsilon_{\bmsub k\lambda}^*\cdot \bm
j_{12}(\bm x),
\eeq
where $\alpha=e^2/4\pi\epsilon_0$ is the fine-structure constant.
If the wavelength of the radiation is large compared to the size
$a$ of the system (i.e. $\omega\ll\Lambda=a^{-1}$) the main
contribution to the integral (\ref{eq:intel1}) comes from small
values of $r=|\bm x|$ ($\omega r\ll 1$). Expanding the exponential
($\bm k=\bm n\omega$)
\beq
\exp(-i\bm k\cdot\bm x)\equiv\exp(-i\omega\bm n\cdot\bm x)=
1-i\omega (\bm n\cdot\bm x)+\frac{(-i\omega)^2}{2!}(\bm n\cdot\bm x)^2+...
\eeq
and integrating term by term one obtains the asymptotic series
\andy{qseries}
\beq
\phi_{\bmsub k\lambda}\sim\sqrt{\frac{2\pi}{V}}\sqrt{\frac{\alpha}{\omega}}
\sum_{s=0}^\infty q_{\bmsub n\lambda}^{(s)}\omega^s,
\label{eq:qseries}
\eeq
where
\andy{a4}
\beq
q_{\bmsub n\lambda}^{(s)}\equiv\frac{(-i)^s}{s!}\int d^3x\;\bm\epsilon_
{\bmsub n\lambda}^*\cdot \bm j_{12}(\bm x)(\bm n\cdot\bm x)^s
\label{eq:a4}
\eeq
($\bm\epsilon_{\bmsub k\lambda}=\bm\epsilon_{\bmsub n\lambda}$
depends only on the direction of $\bm k$). Notice that we explicitly
wrote every $\omega$-dependence and that $q^{(s)}$ does not depend on
$\omega$. Observe that $q^{(0)}$ corresponds to electric-dipole
transitions E1, $q^{(1)}$ to electric quadrupole E2 and/or magnetic
dipole transitions M1, and so on. Hence $s=j-\lambda$, where
$\lambda=0$ ($\lambda=1$) stands for magnetic (electric) transition
Mj (Ej). Since the dominant contribution to the integral in
(\ref{eq:a4}) comes from a region of size $a$ and the current
$j_{12}$ is essentially $\omega_0/a^2$, we get
\beq
q^{(s)}_{\bmsub n\lambda}\propto\omega_0 a^{s+1}, \qquad
\omega_0\equiv E_2-E_1 .
\eeq
If $\omega a\ll 1$ the dominant term in the series
(\ref{eq:qseries}) is the first nonvanishing one, namely
\andy{asympt}
\beq
\phi_{\bmsub k\lambda}\sim q^{(r)}_{\bmsub n\lambda}\omega^r
\propto (\omega_0 a)(\omega a)^r ,
\label{eq:asympt}
\eeq
for some $s=r$. In the continuum limit one gets
\barr
\sum_{\bmsub k,\lambda}|\phi_{\bmsub k\lambda}|^2\longrightarrow
\frac{V}{(2\pi)^3}\sum_\lambda\int d^3k|\phi_{\bmsub k\lambda}|^2
&=&\int_0^\infty
d\omega\;\omega^2\frac{V}{(2\pi)^3}\sum_\lambda\int d\Omega
|\phi_{\bmsub k\lambda}|^2 \nonumber\\ &=& \int_0^\infty
d\omega\;g^2\omega_0\chi^2(\omega),
\earr
where we have defined
\andy{chi1}
\beq\label{eq:chi1}
g^2\omega_0\chi^2(\omega)\equiv
\lim_{V\rightarrow\infty}\frac{\omega^2 V}{(2\pi)^3}\sum_\lambda\int d \Omega
|\phi_{\bmsub k\lambda}|^2,
\eeq
as in (\ref{eq:chi}). From (\ref{eq:asympt}) we obtain
\beq
|\phi_{\bmsub k\lambda}|^2=\frac{2\pi}{V}\frac{\alpha}{\omega}
|\sum_{r=0}^\infty q_{\bmsub n\lambda}^{(r)}\omega^r|^2
\sim \frac{2\pi}{V}\alpha|q_{\bmsub n\lambda}^{(r)}|^2\omega^{2r-1}
\eeq
and therefore
\beq
g^2\omega_0\chi^2(\omega)\sim
\frac{\alpha}{(2\pi)^2}\left(\sum_\lambda\int d\Omega|q_{\bmsub
n\lambda}^{(r)}|^2\right)\omega^{2r+1}
\propto [\alpha (\omega_0 a)^{2r+2}]\omega_0
\left(\frac{\omega}{\omega_0}\right)^{2r+1} .
\eeq
Remembering that $2r+1=2j-2\lambda+1=2j\mp1$, we obtain the first
equation in (\ref{eq:chiprop}) and Eq.\ (\ref{eq:g2}).

On the other hand, if the wavelength is much smaller than $a$ (i.e.
$\omega\gg\Lambda$), we first rewrite (\ref{eq:intel1}) in the
following form
\andy{intel2}
\beq \label{eq:intel2}
\phi_{\bmsub k\lambda}=\sqrt{\frac{2\pi\alpha}{V\omega}}
\int d^3x\;e^{-i\omega\bmsub n\cdot\bmsub x}
\bm\epsilon_{\bmsub k\lambda}^*\cdot \bm j_{12}(\bm x)
=\sqrt{\frac{2\pi\alpha}{V\omega}}
\int dx_\parallel\;e^{-i\omega x_\parallel}
j_{\bmsub n \lambda,12}(x_\parallel),
\eeq
where
\beq
j_{\bmsub n \lambda,12}(x_\parallel)\equiv
\int d^2x_{\perp}\;
\bm\epsilon_{\bmsub n\lambda}^*\cdot \bm j_{12}(\bm x)
\eeq
and $\bm x\equiv x_\parallel \bm n+\bm x_\perp$. According to the
Riemann-Lesbegue lemma, the integral in (\ref{eq:intel2}) vanishes
in the $\omega\rightarrow\infty$ limit. In particular, if
$j_{\bmsub n \lambda,12}(x_\parallel)$ is $N$ times differentiable,
integrating by parts we get
\andy{largeom}
\beq \label{eq:largom}
\phi_{\bmsub k\lambda}=\sqrt{\frac{2\pi\alpha}{V\omega}}
\frac{1}{(i\omega)^N} \int dx\;e^{-i\omega x}
\frac{d^N}{dx^N}j_{\bmsub n \lambda,12}(x)
\eeq
and we can write
\beq
\phi_{\bmsub k\lambda}=o(\omega^{-N-1/2}), \quad(\omega\gg\Lambda)
\eeq
which yields the large $\omega$ behavior of the second equation in
(\ref{eq:chiprop}). It goes without saying that if $j_{\bmsub n
\lambda,12}(x_\parallel)$ is an analytic function, then
$\phi_{\bmsub k\lambda} \to 0$ more rapidly than any power. The
second equation in (\ref{eq:chiprop}) is therefore a conservative
estimate.

\addtocounter{section}{1}
\setcounter{equation}{0}
\section*{Appendix B}
\label{sec-appB}
\andy{appB}

In this Appendix we shall analyze the spectrum of the emitted
photons. We start by substituting (\ref{xsexp}) into (\ref{eq:ys})
and (\ref{eq:zs}), to obtain
\beq
\wtilde y_{\bmsub k\lambda}(s)=\frac{-i\phi_{\bmsub k\lambda}(s+i\omega_k)}
{(s+i\omega_k)^2+B^2}\;\frac{1}{s-s_{\rm pole}},
\eeq
\beq
\wtilde z_{\bmsub k\lambda}(s)=-\frac{\sqrt{\bar N_0}\Phi^*_{\bmsub k_0\lambda_0}
\phi_{\bmsub k\lambda}}
{(s+i\omega_k)^2+B^2}\;\frac{1}{s-s_{\rm pole}}.
\eeq
Closing the Bromwich path with a semicircle in the half plane
$\mbox{Re} s<0$, we get
\barr
y_{\bmsub k\lambda}(t) &=& \frac{1}{2\pi i}\int_{\Gamma}ds\; e^{ts}\wtilde
y_{\bmsub k\lambda}(s), \nonumber \\
z_{\bmsub k\lambda}(t) &=& \frac{1}{2\pi i}\int_{\Gamma}ds\; e^{ts}\wtilde
z_{\bmsub k\lambda}(s),
\earr
which can be evaluated by summing over the integrand residues. The
quantity $|y_{\bmsub k\lambda}(t)|^2 (|z_{\bmsub k\lambda}(t)|^2)$
represents the probability that, at time $t$, the transition
$2\rightarrow1$ ($2 \rightarrow 1 \rightarrow 3$) has taken place.
When $t\rightarrow\infty$, the contribution of $s_{\rm pole}$ (that
has a finite negative real part) is exponentially damped. This
leaves only the contributions of the poles in $-i(\omega_k\pm B)$.

We look first at the case $B=0$ (laser off). One gets
($z_{\bmsub k\lambda} =0,\; \forall t$)
\beq
|y_{\bmsub k\lambda}(+\infty)|^2=\frac{|\phi_{\bmsub k\lambda}|^2}
{(\omega_k-\bar\omega_0)^2+\gamma^2/4}
\eeq
and, in the continuum limit (\ref{eq:chi}), the probability to emit
a photon in the frequency range ($\omega, \omega+d\omega)$ reads
\beq
dP_{B=0}= g^2 \omega_0
\chi^2(\omega)f_L(\omega-\bar\omega_0;\gamma)d\omega,
\eeq
where $f_L$ is the Lorentzian profile
\andy{lorentz}
\beq\label{eq:lorentz}
f_L(\omega;\gamma)=\frac{1} {\omega^2+\gamma^2/4}.
\eeq
This is Eq.\ (\ref{eq:Lprof}) of the text. The quantity $P$ must be
normalized to unity: imposing this condition one gets the Fermi
golden rule (\ref{eq:Fgr}).

On the other hand, when $B\neq0$, the total emission probability is
given by the sum
\beq
|y_{\bmsub k\lambda}(\infty)|^2+|z_{\bmsub k\lambda}(\infty)|^2
\eeq
and it is straightforward to derive the following expressions
($\nu_k=\omega_k-\bar\omega_0$ and we write for simplicity
$\gamma(B)=\gamma$)
\andy{twoll}
\barr
|y_{\bmsub k\lambda}(\infty)|^2 &=& \frac{|\phi_{\bmsub k\lambda}|^2}
{\left|\left(\nu_k+i\frac{\gamma}{2}\right)^2-B^2\right|^2}\left[\left(\nu_
k^2+\frac{\gamma^2}{4}\right)\cos^2(Bt)+B^2\sin^2(Bt)+\frac{\gamma
B}{2}\sin(2Bt)\right],
\nonumber\\
|z_{\bmsub k\lambda}(\infty)|^2 &=& \frac{|\phi_{\bmsub k\lambda}|^2}
{\left|\left(\nu_k+i\frac{\gamma}{2}\right)^2-B^2\right|^2}\left[\left(\nu_
k^2+\frac{\gamma^2}{4}\right)\sin^2(Bt)+B^2\cos^2(Bt)-\frac{\gamma
B}{2}\sin(2Bt)\right],\nonumber\\
\label{eq:twoll}
\earr
which yield
\beq
|y_{\bmsub k\lambda}(\infty)|^2+|z_{\bmsub k\lambda}(\infty)|^2=
\frac{|\phi_{\bmsub k\lambda}|^2}
{\left|\left(\nu_k+i\frac{\gamma}{2}\right)^2-B^2\right|^2}\left(\nu_
k^2+\frac{\gamma^2}{4}+B^2\right).
\eeq
Therefore, in the continuum limit, we can write
\beq
dP_B= g^2 \omega_0 \chi^2(\omega)
\frac{(\omega-\bar\omega_0)^2
+\frac{\gamma^2}{4}+B^2}
{\left[(\omega-\bar\omega_0-B)^2+\frac{\gamma^2}{4}\right]
\left[(\omega-\bar\omega_0+B)^2+\frac{\gamma^2}{4}\right]}d\omega.
\eeq
This formula can be rewritten in the following form
\beq
dP_B=g^2 \omega_0 \chi^2(\omega)
\frac{1}{2}
\left[f_L(\omega-\bar\omega_0-B;\gamma)+
f_L(\omega-\bar\omega_0+B;\gamma)\right]d\omega.
\eeq
This is Eq.\ (\ref{eq:LLprof}) of the text. We see that the
emission probability is the sum of two Lorentzians, centered in
$\bar\omega_0-B$ and $\bar\omega_0+B$ and weighted by $g^2 \omega_0
\chi^2(\omega)$. This result is in agreement with that obtained
in Refs.\ \cite{ZNS,LSS97}. Incidentally, we notice that the value
(\ref{eq:gamma(B)}) of $\gamma(B)$ can be readily estimated by
imposing the normalization of the emission probability
\andy{twoL}
\beq
\int dP_B=\int_{0}^{\infty} g^2 \omega_0 \chi^2(\omega)\frac{1}{2}
\left[f_L(\omega-\bar\omega_0-B;\gamma)+
f_L(\omega-\bar\omega_0+B;\gamma)\right]d\omega=1 .
\label{eq:twoL}
\eeq
Performing the integration one obtains ($\gamma\ll\bar \omega_0$,
hence one can integrate over the whole real axis and take
$\chi^2(\omega)$ equal to its value on each Lorentzian peak)
\andy{gamcc}
\beq
1=\int dP_B \approx\frac{1}{2}g^2\omega_0
\left[ \chi^2(\bar\omega_0+B) + \chi^2(\bar\omega_0-B) \right]
\frac{2\pi}{\gamma(B)},
\label{eq:gamcc}
\eeq
which yields Eq.\ (\ref{eq:gamma(B)}) of the text.


\end{document}